\documentclass[final,5p,times,twocolumn,number]{elsarticle}

\usepackage{amssymb}
\usepackage{lipsum}
\usepackage{xcolor}
\usepackage{amsmath}

\usepackage{graphicx}
\usepackage{amsthm}
\usepackage{amsmath}
\usepackage{graphicx,subfigure}
\usepackage{dcolumn}
\usepackage{bm}
\usepackage{slashed}
\usepackage{calligra}
\DeclareMathAlphabet{\mathcalligra}{T1}{calligra}{m}{n}
\DeclareFontShape{T1}{calligra}{m}{n}{<->s*[2.5]callig15}{}
\newcommand\brabar{\scalebox{.4}{(}\raisebox{-1.7pt}{--}\scalebox{.4}{)}} 
\usepackage{stackengine}
\newcommand{\be}{\begin{eqnarray}}
\newcommand{\ee}{\end{eqnarray}}
\newcommand{\bea}{\begin{eqnarray}}
\newcommand{\eea}{\end{eqnarray}}

\biboptions{sort&compress}

\begin{document}

\begin{frontmatter}
\title{Excitation of the Glashow resonance without neutrino beams}

\author{Ibragim Alikhanov}
\ead{ialspbu@gmail.com}
\address{North-Caucasus Center for Mathematical Research,\\ North-Caucasus Federal University, Stavropol 355017, Russia}

\begin{abstract}
The $s$-channel process $\bar\nu_ee^-\rightarrow W^-\text{(on-shell)}$ is now referred to as the Glashow resonance and being searched for at kilometer-scale neutrino ice/water detectors like IceCube, Baikal-GVD or  KM3NeT. After over a decade of observations, IceCube has recorded only a few  neutrino events with energies of interest such that an independent confirmation of the existence of this resonant interaction would be of great importance for testing the Standard Model. One might therefore ask: are there reactions with the Glashow resonance that would not necessitate having initial (anti)neutrino beams? This article suggests a surprisingly affirmative answer to the question -- namely, that the process may proceed in  electron--positron collisions at accelerator energies, occurring as~$e^+e^-{\rightarrow}\,W^-\rho(770)^+$. Although the resonance appears somewhat disguised, the underlying physics is transparent, quite resembling the well known radiative return: emission of~$\rho^+$ from the initial state converts the incident~$e^+$~into~$\bar\nu_e$. Likewise, the CP conjugate channel, $\nu_e e^+\rightarrow W^+$, takes the form $e^+e^-\rightarrow W^+\rho(770)^-$. Similar reactions with muons and other hadrons are also possible. From this viewpoint, future high-luminosity lepton colliders seem to be promising for excitation of the Glashow resonance in laboratory conditions.
\end{abstract}


\end{frontmatter}

\section{Introduction}
In order to explore production of massive states in neutrino-induced reactions, one should have at disposal  high energy neutrino beams. On the other hand, in the
laboratory reference frame,  energy-momentum conservation results in high longitudinal momenta of the final particles to be accurately evaluated for reconstructing the energy of an event. For this reason, the volume of a detector should be large enough to allow the particles to leave an interpretable signal in the bulk.  Meeting or arranging these conditions is always a challenge for physicists and engineers~\cite{detector_rev}. This is particularly the case of searches for the Glashow resonance -- annihilation of an electron antineutrino with an electron into an on-shell $W^-$ boson, $\bar\nu_ee^-\rightarrow W^-$. The existence of this resonant process had been predicted by Sheldon Glashow in 1959~\cite{glashow_res} and still needs to be experimentally studied. 

To produce a state as heavy as $W^-$ ($m_W\eqsim$~80.3~GeV) in scattering with electrons at rest, the incident neutrino energy must be very high, about $6.3\times10^{15}$~eV (6.3~PeV)~\cite{gazizov}. The only available source of such energetic particles  is cosmic rays. However, the relatively low intensity of the astrophysical neutrino beam requires significant observation times, even at kilometer-scale telescopes like IceCube~\cite{icecube_review}, Baikal-GVD~\cite{baikal} or  KM3NeT~\cite{km3net}. Thus, after over a decade of searches, IceCube has recorded only a few neutrino events with PeV energies~\cite{icecube_nature:2021,over_pev_icecube1,over_pev_icecube2}. Of these, one shower appeared in the vicinity of the resonance, having an energy~of~$6.05\pm0.72$~PeV~\cite{icecube_nature:2021,Distefano_nature:2021}. The candidate events should not, of course, necessarily fall into the energy bins close to the resonance peak. Being directly related to the flavor composition and the spectrum of ultra-high energy cosmic neutrinos~\cite{gr_events1,gr_events2},   the events in principle can have energies of about~1--2~PeV~\cite{res_lower1,res_lower2,res_lower3}. It should also be mentioned that KM3NeT has recently reported observation of a cosmic neutrino with an energy of~$120{+110\atop -60}$~PeV~\cite{KM3NeT:2025npi}, which is, however, far above the resonance. Various implications of the detection of the Glashow resonance as well as the expected signatures  have been extensively studied, e.g., in ~\cite{gr_diff1,gr_diff2,gr_diff3,gr_diff4,gr_diff5,gr_diff6,gr_diff7,gr_diff8}. 

An independent experimental confirmation of this long-standing prediction  would be of great importance for testing the Standard Model.
One might therefore ask: are there reactions with the Glashow resonance that would not necessitate having initial (anti)neutrino beams? We suggest a surprisingly affirmative answer to this question -- namely, that the process may proceed in  electron--positron collisions at accelerator energies, occurring, for instance, as $e^+e^-\rightarrow W^-\rho(770)^+$. Although the resonance appears somewhat disguised, the underlying physics is transparent, quite resembling the well known radiative return: emission of~$\rho^+$ from the initial state converts the incident~$e^+$ into~$\bar\nu_e$. Likewise, the CP conjugate channel, $\nu_e e^+\rightarrow W^+$, takes the form $e^+e^-\rightarrow W^+\rho(770)^-$. 
From this viewpoint, future high-luminosity lepton colliders seem to be promising for excitation of the Glashow resonance. This contrasts with the settled opinion that the principle impediment to observing the resonance  in laboratories is the limited energies attainable by  terrestrial (i.e.~human-made) accelerators. The suggested reactions might occur  already at total energies of order 100~GeV in the center-of-mass (cms) frame. 
The possibility of accessing the resonance with charged lepton beams  was also addressed in~\cite{effective_neitrinos1,effective_neitrinos2}.

\section{Initial state photon emission \label{sec:isr}}
To lay the groundwork for the subsequent analysis, in this section we outline a well known effect in quantum electrodynamics (QED) --  initial state radiation (ISR).   

As a specific example, let us consider electron--positron annihilation into the $Z$ boson, $e^+e^-\rightarrow Z$. This annihilation is frequently accompanied by photon emission,
\be
\label{eq:zgamma}
e^+e^-\rightarrow Z\gamma,
\ee  
which distorts the $Z$ line shape~\cite{altarelli:89}.
The underlying physics becomes especially transparent in the framework of the equivalent particle approximation~\cite{zerwas:1975}. Even if the initial total energy of the collision is higher than the $Z$ boson mass, $s>m_Z^2$, the energy excess is carried away by the emitted photon, thus turning the $e^+e^-$ pair to the resonance pole~\cite{fadin:1968}. This gives rise to the so called radiative tail -- a widening of the resonance to the right of its peak. The possibility of  ``tuning'' the $e^+e^-$ collision energy by ISR is the basis of the radiative return~\cite{radiative_return}. 

\begin{figure}[t]
\centering
\includegraphics[width=0.43\textwidth]{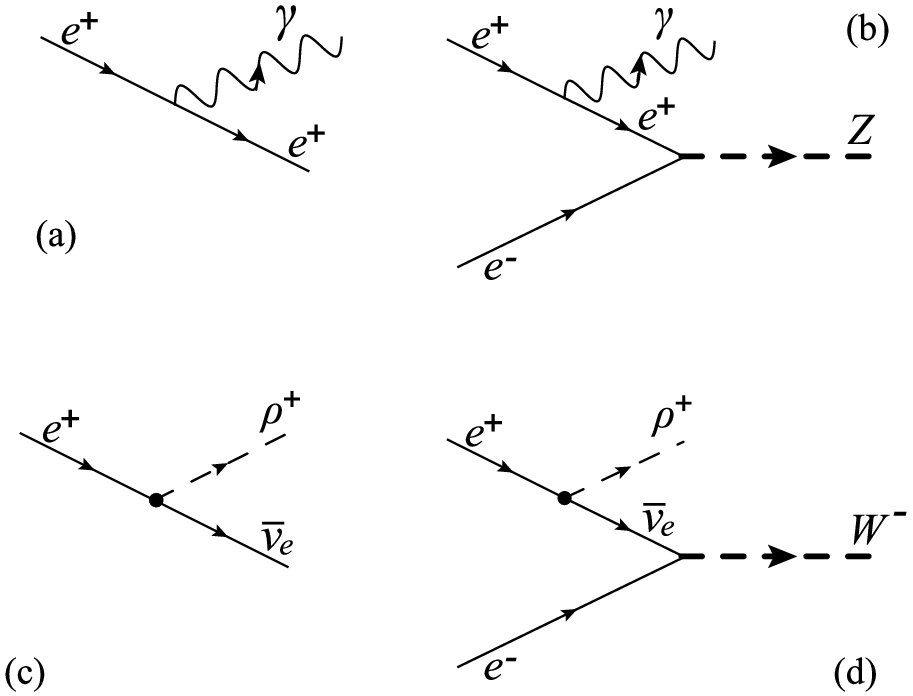} \\
\caption{Diagrams illustrating: (a) the QED photon coupling to the positron current; (b) annihilation of the $e^+e^-$ pair into $Z$ boson accompanied by ISR, $e^+e^-\rightarrow Z\gamma$; (c) the effective coupling of $\rho^+$ meson to the electroweak charged current;  (d) annihilation of the $\bar\nu_ee^-$ pair into $W^-$ boson accompanied by initial state $\rho^+$ emission, $e^+e^-\rightarrow W^-\rho^+$. The arrows sketch the spatial momentum flows.
}
\label{fig:isr_rho}
\end{figure}

Using the language of the quark--parton model~\cite{parton_model, qpm_model}, one can derive the probability density of finding a positron with energy fraction $x$ in the parent positron~\cite{zerwas:1975}. In the leading order it is of the form
\be\label{eq:distr}
F_{e/e}(x,Q^2)=\frac{e^2}{8\pi^2}\frac{1+x^2}{1-x}\ln\left(\frac{Q^2}{m_e^2}\right),
\ee
where $e=\sqrt{4\pi\alpha}$ is the elementary electric charge, $Q^2$~is four-momentum transfer squared, $m_e$ is the electron mass. This result is a direct consequence of the QED photon coupling to the positron current with the matrix element
\be
\label{qed_coupling}
 \mathcal{M}_\text{QED}=e \left [\bar v_e\gamma^\sigma v_e\right] \varepsilon^*_\sigma.  
\ee
The corresponding diagram is shown in Fig.~\ref{fig:isr_rho}(a). By CP symmetry, exactly the same distribution holds for the electron. Details of derivation of~\eqref{eq:distr} can be found in textbooks,  e.g. in~\cite{peskin}.

The cross section for process \eqref{eq:zgamma} is represented as the following integral~\cite{radiative_return}:
\be
\label{cross_zgamma}
\sigma_{ee\rightarrow Z\gamma}(s)=2\int F_{e/e}(x,s)\,\sigma_{ee\rightarrow Z}(xs)\,\mathrm{d}x,  
\ee
where $\sigma_{ee\rightarrow Z}(s)$ is the cross section for the annihilation~$e^+e^-{\rightarrow}\,Z$, the factor of 2 takes into account the possibility that either initial lepton can radiate the photon. Note that the simple kinematics of the process gives $Q^2=s$ in the high energy limit. It is crucial that not only the photon but also the leptons can be regarded massless in comparison with the heavy $Z$ boson. An additional insight into the described mechanism may be provided by the diagram in Fig.~\ref{fig:isr_rho}(b).

Using the narrow width approximation, $\sigma_{ee\rightarrow Z}(s)=12\pi^2\Gamma_{Z\rightarrow ee}\delta(s-m_Z^2)/m_Z$, one can check~\cite{epa_threshold} that the integration~\eqref{cross_zgamma} yields exactly the leading order result~\cite{exact_Zgamma} of straightforward electroweak calculations of the cross section. In the latter expression, $\Gamma_{Z\rightarrow ee}$ is the width of the $Z\rightarrow e^+e^-$ decay channel, $\delta(x)$ is the Dirac delta function.

The representation considered allows us to see analytically the presence of the resonance. Even though there are two particles in the final state, $Z\gamma$, instead of a single $Z$, we still deal with the resonant production of the boson through $e^+e^-\rightarrow Z$. Experimentally, the mechanism manifests itself as a distortion of the resonance line shape~\cite{altarelli:89}.

\section{Initial state meson emission \label{sec:initial_rho}}
The Standard Model admits effective couplings of charged vector mesons to the leptonic currents. They had been known long ago and even entered textbooks~\cite{okun}. Such an interaction of the $\rho(770)^+$ meson is represented graphically in  Fig.~\ref{fig:isr_rho}(c). The coupling opens, for example, channels for the mesons to decay into lepton pairs~$\stackon[.1pt]{$\nu$}{\brabar}_{\hskip -1mm l} \,l^\mp$~\cite{meson_decay2}. Inverse processes, with the production of $\rho(770)^\pm$ in lepton--lepton scattering, are also allowed~\cite{meson_decay4,meson_decay5}. Henceforth, the symbol $\rho^\pm$ will stand for the meson state $\rho(770)^\pm$.

In addition to the processes mentioned, the effective coupling~$e\nu\rho$ also gives rise to the following reaction:
\be
\label{eq:main_reaction}
 e^+e^-\rightarrow W^-\rho^+,
\ee 
as illustrated by  Fig.~\ref{fig:isr_rho}(d).
 The apparent similarity with the diagram~in Fig.~\ref{fig:isr_rho}(b)  suggests that in~\eqref{eq:main_reaction} we may have a mechanism resembling ISR. As will be seen, this interpretation is tenable.

Compare the diagrams in Figs.~\ref{fig:isr_rho}(a) and \ref{fig:isr_rho}(c). From the kinematical point of view, both subprocesses are equivalent at high energies since it becomes fair to neglect the meson mass. By analogy with~\eqref{eq:distr}, we associate a distribution of electron antineutrinos with the positron.  
The vertex~$e\nu\rho$ in Fig.~\ref{fig:isr_rho}(c) gives a matrix element (see, e.g.,~\cite{meson_decay2}) 
\be
\label{eff_coupling}
 \mathcal{M}_{\text{effective}}=\frac{G_F}{\sqrt{2}}V_{ud}\,f_\rho\, m_\rho\left[\bar v_e\gamma^\sigma(1-\gamma^5) v_{\nu}\right]\epsilon^{*}_\sigma,  
\ee
where $G_F$ is the Fermi coupling constant, $V_{ud}$ is the Cabibbo--Kobayashi--Maskawa (CKM) matrix
element between the $u$ and $d$ quarks, $f_\rho$ is the decay constant of the meson parametrizing the hadronic part of the amplitude and $m_\rho$ is the mass of the meson. The polarization vector $\epsilon_\sigma$ satisfies the standard summation $\sum_{\text{polarizations}}\epsilon_\lambda\epsilon^{*}_\sigma=-\left(g_{\lambda\sigma}-p_{\lambda}p_{\sigma}/m^2_\rho\right)$  
with $p_{\lambda,\sigma}$ being the four-momentum of the meson. 
Simple calculations (essentially the same as those in QED described above) yield the probability density of finding an electron antineutrino with energy fraction~$x$ in the parent positron:
\be\label{eq:distr_nu}
F^{(\rho)}_{\nu/e}(x,Q^2)=\frac{\left(G_\rho G_F\right)^2}{8\pi^2}\frac{1+x^2}{1-x}\ln\left(\frac{Q^2}{m_\rho^2}\right).
\ee
Here we have written $G_\rho$ as a shorthand for the product $|V_{ud}|f_\rho m_\rho$.  The superscript $(\rho)$ on $F$ labels the meson whose emission yields the distribution. Note that $f_\rho$ has units of energy, so that $G_\rho G_F$ is dimensionless. CP symmetry and lepton universality are assumed throughout. 

The similarity between~\eqref{eq:distr_nu}  and~\eqref{eq:distr} is evident, as could have been anticipated from the kinematics. We therefore suggest an interpretation analogous to the radiative return~$-$~namely, that reaction~\eqref{eq:main_reaction} may proceed via initial state meson emission.    
The crucial subtlety is that the emission of $\rho^+$ not merely carries away the energy excess but converts the incident $e^+$ into~$\bar\nu_e$. The latter annihilates with the electron into the final~$W^-$. In other words, the Glashow resonance might occur as $e^+e^-\rightarrow W^-\rho^+$. 
As in~\eqref{cross_zgamma}, the presence of the resonance becomes analytically explicit in the partonic representation of the corresponding cross section: 
\be
\label{eq:cross_Wrho}
\sigma_{ee\rightarrow W\rho}(s)=\int F^{(\rho)}_{\nu/e}(x,s)\,\sigma_{\nu e\rightarrow W}(xs)\,\mathrm{d}x, 
\ee
where $\sigma_{\nu e\rightarrow W}(s)$ is the cross section for $\bar\nu_ee^-\rightarrow W^-$.
We again have taken into account that here $Q^2=s$ in the high energy limit. The integral reminds us of the description of a Drell--Yan process~\cite{Drell:1970wh} in which one of the two projectiles remains on-shell (electron) and annihilates with partons (antineutrinos) distributed with the density $F$ in the other projectile (positron). The factorization in~\eqref{eq:cross_Wrho} is justified as long as $s\gg m^2_\rho$.


Within the narrow width approximation, which gives $\sigma_{\nu e\rightarrow W}(s)=\sqrt{8\pi^2} G_Fm_W^2\delta(s-m_W^2)$, we find that
\be
\label{eq:cross_Wrho1}
\sigma_{ee\rightarrow W\rho}(s)=\frac{G^2_\rho G_F^3}{\sqrt{8\pi^2}}\frac{r+r^3}{1-r}\ln\left(\frac{s}{m_\rho^2}\right),
\ee
where $r=m_W^2/s$. It may be instructive to write the result also as $\sigma_{ee\rightarrow W\rho}(s)=\sqrt{8\pi^2} G_F\,rF^{(\rho)}_{\nu/e}(r,s).$
The cross section turns out to be proportional to the antineutrino distribution in the positron. Projecting out  a parton distribution in a cross section is a general indicator of excitation of a narrow resonance~\cite{zerwas:1975}. In our case, the ratio of the total width of $W^-$ to its mass is $\eqsim1/40$ such that the assumption about the relative narrowness of the Glashow resonance is acceptable. The same situation, by the way, occurs in $e^+e^-\rightarrow Z\gamma$. One can verify that its leading order cross section~\cite{exact_Zgamma} is a projection of the QED electron distribution $F_{e/e}(x,s)$.

As may be already obvious, the discussed mechanism  is realizable not only with $\rho^\pm$, but also with a relatively wide set of scalar and vector particles (not necessarily on-shell) provided: (i) they  couple to the electroweak currents; (ii) their squared four-momenta are much smaller compared to those of the produced final states. In general, the process will have the form: $e^+e^-\rightarrow W^-h^+$, where $h^+$ is any allowed state satisfying the above conditions. For example, from the kinematical point of view, a favourable reaction to excite the Glashow resonance would be $e^+e^-\rightarrow W^-\pi^+$ due to the smallness of the pion  mass, but its probability is  helicity-suppressed. This is the reason we have chosen the $\rho$ meson to illustrate the resonant mechanism. It is a vector particle, free from the helicity-suppression and, in addition, relatively light. On the other hand, dealing with the specific familiar charged meson,  our discussion is not abstract. Another promising channel may be $e^+e^-\rightarrow W^-D^{*+}_s$, where $D^{*+}_s$ is  the meson of mass $m_{D^{*}_s}=2112.2$~MeV with explicitly nonzero quantum numbers of charm and strangeness, $C = S = 1$~\cite{pdg}. The width and decay modes of $D^{*\pm}_s$ are consistent with the spin-parity $J^P=1^-$. Its coupling to the leptonic weak current is favored by the CKM element~$|V_{cs}|$.

\section{CP symmetry and lepton universality}

There is a lack of attention given to the CP~conjugate of the Glashow resonance ($\nu_e e^+\rightarrow~W^+$) and its muonic counterpart ($\bar\nu_\mu \mu^-\rightarrow~W^-$) in the literature. Meanwhile, these channels are also crucial predictions of the Standard Model still requiring experimental verification. The latter would furnish important support for CP symmetry and lepton universality in the neutrino sector. The continued omission of these channels can be explained by the challenging problem of designing experiments on high momentum-transfer neutrino--positron and neutrino--muon collisions.
Here, we point out that the parton model offers a framework for solving this problem. 
For example, it has been known that with each electrically charged particle one can associate distributions of charged leptons, $e^\pm$, $\mu^\pm$ and, in some cases, even $\tau^\pm$~\cite{zerwas:1975}. To put another way,  with a certain probability depending on kinematical conditions a particle is able to manifest itself as a positron, a muon or a tau lepton. The distributions are the QED analogues of those for the quarks and gluons such that accurately computable by well-developed methods. One thus can exploit atomic nuclei as sources of positrons and muons to get access to the channels like $\stackon[.1pt]{$\nu$}{\brabar}_{\hskip -0.8mm\l} e^+\rightarrow X$ and $\stackon[.1pt]{$\nu$}{\brabar}_{\hskip -0.8mm\l}\mu^\pm\rightarrow X'$ producing massive final states $X$~\cite{alikhanov_gr1,alikhanov_gr2}. This is closely related to the ability of photons to split into $l^+l^-$ pairs~\cite{zerwas:1975,photon_splitting1,photon_splitting2} in interactions with neutrinos~\cite{alikhanov_photon1,alikhanov_photon2}.

\begin{figure}[t]
\centering
\includegraphics[width=0.48\textwidth]{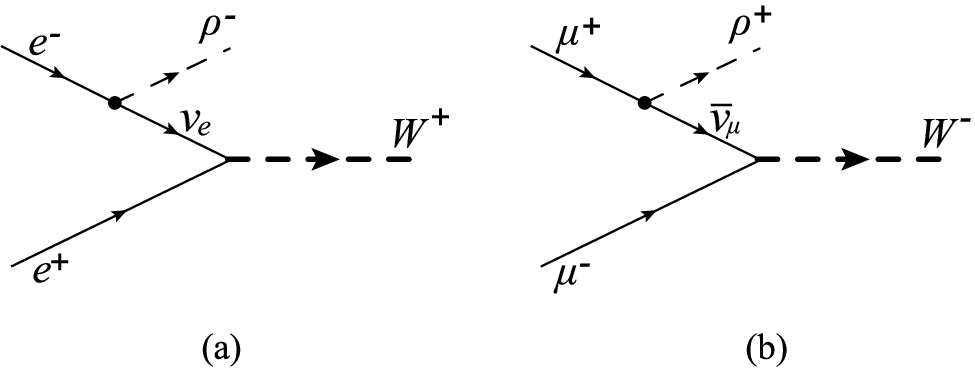} \\
\caption{Diagrams illustrating: (a) excitation of the CP conjugate of the Glashow resonance accompanied by initial state $\rho^-$ meson emission,~$e^+e^-{\rightarrow}\,W^+\rho^-$; (b) excitation of the muonic counterpart of the Glashow resonance accompanied by initial state $\rho^+$ meson emission, $\mu^+\mu^-\rightarrow W^-\rho^+$. The arrows sketch the spatial momentum flows. 
}
\label{fig:cp_conjugate}
\end{figure}

On the other hand, as previously discussed, a neutrino distribution can also be linked to the electron. By lepton universality, the same must hold for muons as well. This could allow for testing both the CP conjugate of the Glashow resonance and its muonic counterpart via the reactions $e^+e^-\rightarrow W^+\rho^-$, $\mu^+\mu^-\rightarrow W^\mp\rho^\pm$ (see  Fig.~\ref{fig:cp_conjugate}). In the Standard Model, their cross sections coincide at high energies, being given  by~\eqref{eq:cross_Wrho1}.

\section{Nonresonant background}
By nonresonant background we mean those channels in $e^+e^-$  ($\mu^+\mu^-$) scattering that produce exactly the same final states as above, $W^\mp\rho^\pm$, without, however,  exciting the Glashow resonance. 

\begin{figure}[t]
\centering
\includegraphics[width=0.48\textwidth]{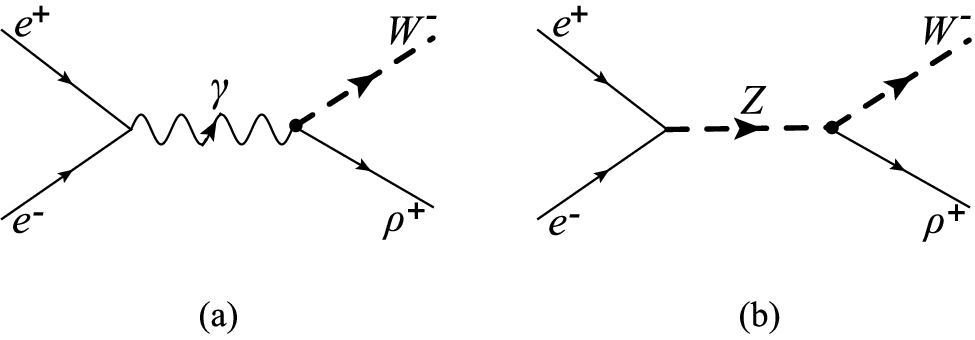} \\
\caption{Leading-order Feynman diagrams for nonresonant background: (a)~with $s$-channel $\gamma$ exchange; (b) with $s$-channel $Z$ boson exchange.  The arrows sketch the spatial momentum flows.  
}
\label{fig:background}
\end{figure}

The possible leading order background can in general be represented by two $s$-channel  Feynman diagrams with $\gamma$ and $Z$ exchanges as shown in Fig.~\ref{fig:background}. For completeness, one could also add a similar graph with an intermediate Higgs boson, but we ignore it. In principle, the vertices $\gamma W\rho$ and $ZW\rho$ may arise in effective theories like, e.g., vector meson dominance extended to weak interactions~\cite{vmd_weak}.
These diagrams would yield terms in a cross section behaving like $\sim m_W^2/s^2$ as $s$ grows, and without the logarithmic enhancement. As a result, the related contribution will decrease much faster with energy than ~$1/s\ln(s/m^2_\rho)$ following from~\eqref{eq:cross_Wrho1}. Another independent reason of suppression of the background is due to the couplings $\gamma W\rho$ and $ZW\rho$. The latter are constrained by the non-observation of the decay modes $W^\pm\rightarrow\gamma\rho^\pm$ and $Z\rightarrow W^\pm\rho^\mp$. Note that at the $Z$-pole the contribution of the diagram in Fig.~\ref{fig:background}(b) suffers, in addition, a phase-space suppression due to the mass difference $m_Z-m_W\sim 10$~GeV. The current 95\%~CL upper bound on the branching ratio $\Gamma(Z\rightarrow W^\pm\rho^\mp)/\Gamma(Z\rightarrow\text{all})$ is smaller than $8.3\times10^{-5}$~\cite{pdg} and can further drop to $4.0\times10^{-10}$~\cite{Grossman:2015cak}.
Thus, the nonresonant channels seem to play a minor role~(if~any) in the production of final states consisting of a single $W$ boson plus a vector meson.

\section{Numerical results}
There have been various proposal for future high-luminosity lepton colliders~\cite{collider_review}. Among other things, their goal is to explore physics at the $Z$ ($\sqrt{s}=91$ GeV) and  Higgs boson ($\sqrt{s}=125$ GeV) poles. These energies are also sufficient for the resonant reactions under consideration to proceed, for instance, as $e^+e^-\rightarrow W^\mp\rho^\pm$ and $e^+e^-\rightarrow W^\mp D^{*\pm}_s$. 
Table~\ref{tab:table1} summarizes our estimates of the associated number of events that could occur at  collider experiments assuming integrated luminosity of 10~ab$^{-1}$.
The cross sections are calculated by  formula~\eqref{eq:cross_Wrho1}, with the obvious replacements $|V_{ud}|\rightarrow|V_{cs}|$, $f_\rho\rightarrow f_{D^*_s}$, $m_\rho\rightarrow m_{D^{*}_s}$ in case of $D^{*\pm}_s$. 
The values of $G_F$, $|V_{ud}|$, $|V_{cs}|$ as well as the masses of $\rho^{\pm}$, $D^{*\pm}_s$ and $m_W$ are taken from the Particle Data Group~\cite{pdg}. To be sure that the adopted approximation is justified we leave out of consideration the $W^\mp D^{*\pm}_s$ channel at $\sqrt{s}=91$~GeV.  
 
The results show that the rate of excitation of the Glashow resonance at future colliders might be higher than that at the currently conducted experiments exploiting natural  neutrino beams. For example, at the $Z$ pole, the run of the collider CEPC~\cite{collider_cepc} in 2 years with two interaction points (IP) would provide integrated luminosity of~100~ab$^{-1}$, while the 4 years run of the FCC-ee, also with 2 IPs~\cite{collider_fccee}, corresponds to~192~ab$^{-1}$.
Instantaneous luminosities at these colliders are foreseen to be of order~$10^{36}$~cm$^{-2}$s$^{-1}$ which could lead to a rate greater than~10~events/year  from~the~$W^\mp\rho^\pm$~channel~only. Taking into account a more complete set of allowed mesons will further increase the number of events expected in a laboratory. For comparison, IceCube observes less than~1~event per year in the resonance energy region~~\cite{icecube_nature:2021,over_pev_icecube1,over_pev_icecube2}.

\begin{table}[h!]
  \begin{center}
    \caption{Number of $e^+e^-\rightarrow W^\mp\rho^\pm$ and $e^+e^-\rightarrow W^\mp D^{* \pm}_s$ events at  collider experiments assuming integrated luminosity of 10~ab$^{-1}$. Two different values of the decay constants are used for each channel. The numbers in square brackets indicate the relevant references.}
    \label{tab:table1}
    \begin{tabular}{c|r|c|c}\hline 
     {}&{Decay const.} & No. of events & No. of events\\
     Channel &$f_{\rho,D^*_s}$ (MeV)& $\sqrt{s}=91$ GeV & $\sqrt{s}=125$ GeV \\
      \hline
     $ee\rightarrow W\rho$\,\,\,\,\, &219  \,\,\,\cite{tab_f219_490} & 2.0 & 0.3\\
     $ee\rightarrow W\rho$\,\,\,\,\, &490   \,\,\,\cite{tab_f219_490} & 10.3 & 1.6\\
     $ee\rightarrow WD^*_s$ &240   \,\,\,\cite{tab_f240} & -- & 2.3\\
     $ee\rightarrow WD^*_s$ &391   \,\,\,\cite{tab_f391}& -- & 6.0\\\hline
    \end{tabular}
  \end{center}
\end{table}

We would like to highlight the important role played by the decay constants in determining the observables. Being nonperturbative, these parameters are not known precisely. Different models give more o less different values (see, e.g.,~\cite{decay_const}).  To stress the fact that the related uncertainties may significantly affect numerical predictions,  we have used two different values of the constants for each channel (see Table~\ref{tab:table1}).  

It is interesting to speculate on the possibility for the parameters to depend on the momentum transfer squared. For example, their growth in high energy reactions would lead to an enhancement of the associated event rate.  Meanwhile, the decay constants used are originally adopted for the description of the decays of mesons with masses~$\sim$1~GeV~\cite{meson_decay2}.

The estimates given in Table~\ref{tab:table1} are equally valid for high luminosity $\mu^+\mu^-$ collides. The latter could excite the muonic counterpart of the Glashow resonance, thus  testing  lepton universality of this  process. Precision measurements at a 125 GeV muon collider with integrated luminosity of 20 ab$^{-1}$ have already been discussed~\cite{mu_collider}.

\section{Conclusions}
Due to the existence of the radiative return, one can measure the cross section for the $s$-channel annihilation~$e^+e^-\rightarrow~X$ in a relatively wide range of final state masses, $m_X$, using only one collider operating at a fixed cms energy~\cite{radiative_return}. As a result, the annihilation is explored in reactions of the form $e^+e^-\rightarrow X\gamma$. Without this mechanism, in order to carry out the same task, one would have to obtain new $e^\pm$~beams  for each value of~$m_X$.     

There are mesons that also  couple to the leptonic currents. From the kinematical point of view, the difference between photon emission and emission of such a meson disappears in the high energy limit.  Therefore, it seems to be reasonable to assume the existence of the meson analogue of the radiative return. If the meson is positively charged, its emission from the initial state not merely carries away a fraction of the collision energy but converts the incident $e^+$ into~$\bar\nu_e$.  In other words, with certain calculable probability the positron interacts as the related antineutrino. 
We have obtained this probability and suggested a mechanism of initial state charged meson emission. This allowed us  to factorize the reaction $e^+e^-\rightarrow W^-\rho^+$ into two processes and thus put forward a Glashow resonance interpretation of the underlying physics. Similar channels with muons and other hadrons have also been considered.  From this viewpoint, future high-luminosity lepton colliders seem to be promising for excitation of the Glashow resonance, its CP~conjugate and its muonic counterpart. Note that the latter two  are widely believed today to be inaccessible in neutrino experiments. 
The reactions might occur  already at cms energies of order 100~GeV.  
The signal will consist of a single $W$ boson plus a meson or a jet (oppositely charged with respect to $W$). If the accompanied meson (jet) is not detected, the process will manifest itself as a single $W$ boson. The energy region of interest, by the way, have  been investigated in the  LEP experiments at CERN~\cite{lep_cern}. It might be therefore interesting to look for the signals among the final states with a single $W$ measured by LEP~\cite{lep_singlew1,lep_singlew}, at least for constraining the parameters.

While our numerical estimates provide a baseline, further analysis will be necessary for more detailed predictions. The integration of the reactions into computational simulation tools, such as  {\sc\small MadGraph5\_aMC@NLO}~\cite{Alwall:2014hca}, PYTHIA~8~\cite{Sjostrand:2007gs} and GEANT4~\cite{GEANT4:2002zbu}, would facilitate a precise evaluation of signal-to-background ratios within various experimental contexts.

In conclusion, we emphasize that the approach developed here is by no means limited to the specific reactions analyzed; it offers a framework for  investigating a wide range of neutrino-initiated $s$ and $t$-channel processes  (including those beyond the Standard Model) in lepton--lepton, lepton--hadron or lepton--ion collisions.

\section*{Acknowledgements}
The author thanks E.~A.~Paschos for valuable comments.   
 He has also profited from conversations or correspondence with  A.~B.~Arbuzov, A.~D. Dolgov, B.~I. Ermolaev and A.~A.~Osipov.
\bibliographystyle{elsarticle-num}

\bibliography{references}

\end{document}